# Design of Current Controller for Two Quadrant DC Motor Drive by Using Model Order Reduction Technique

K.Ramesh
EEE Department,
Velalar College of Engg. & Tech.,
Erode - 638012, India

K.Ayyar
EEE Department,
Velalar College of Engg. & Tech.,
Erode - 638012, India

Dr.A.Nirmalkumar
EEE Department,
BIT, Sathyamangalam,
India

Dr.G.Gurusamy
EEE Department,
BIT, Sathyamangalam
India

*Abstract* -In this paper, design of current controller for a two quadrant DC motor drive was proposed with the help of model order reduction technique. The calculation of current controller gain with some approximations in the conventional design process is replaced by proposed model order reduction method. The model order reduction technique proposed in this paper gives the better controller gain value for the DC motor drive. The proposed model order reduction method is a mixed method, where the numerator polynomial of reduced order model is obtained by using stability equation method and the denominator polynomial is obtained by using some approximation technique preceded in this paper. The designed controller's responses were simulated with the help of MATLAB to show the validity of the proposed method.

## I. INTRODUCTION

DC motors used in many applications such as steel rolling mills, electric trains and robotic manipulators require current and speed controllers to perform tasks. Major problems in applying a conventional control algorithm in a controller design are the effects of nonlinearity in a DC motor. The nonlinear characteristics such as friction and saturation could degrade the performance of conventional controllers. Many advanced model-based control methods such as variable structure control and model reference adaptive have been developed to reduce these effects. However, the performance of these methods depends on the accuracy of system models and parameters. In this paper current controller of two quadrant DC motor drive is considered. The linear operation of DC motor drive was taken in to account in the design stage of current controller. In conventional design methods, some of simplification processes are considered to design the controller parameter values but where as in this proposed method, the model order reduction technique was introduced for the controller parameter design values.

Both in systems and control engineering and in numerical analysis, a wealth of model order reduction techniques have been developed. Balanced truncation, Krylov subspace methods, proper orthogonal decomposition and other SVD-based methods are just a few classes of methods that have been developed.

The computation of equivalent linear system models of large linear dynamic systems is a topic of considerable practical interest. This interest is motivated by the reduced complexity obtained by reducing the large linear sub-network in a linear (or nonlinear) network. Ideally, linear analysis on these sub-networks is performed by first computing a state space model or equivalent transfer function form, followed by the application suitable analysis method. However, the applicability of this method is limited since typical dynamic systems are represented by very large scale matrices that require specialized large-scale eigen analysis programs and computer resources. To avoid this practical limitation, model-order reduction methods are widely used in the solution of such systems. The basic idea behind model-order reduction is to replace the original system equations with a much smaller state-space or transfer function dimension. In particular, the identified reduced order model frequency characteristics must approximate those of the full order model.

In the analysis of many systems for which the physical laws are well known, one is frequently confronted with problems arising from the high dimensions of descriptive state model, the famous curse of dimensionality. The reduction of such high order systems (also termed as large scale systems) into low order models is one of the important problems in control and system theory system and is considered important in analysis, synthesis and simulation of practical systems. The exact analysis of high order systems is both tedious and costly.

To overcome the stability problem Hutton & Friedland [1] and Appiah [2] gave different methods, called stability based reduction methods which make use of some stability criterion. Other approaches in this direction include the methods such as Shamash [3] and Gutman, Mannerfelt & Molandor [4] which do not make use of any stability criterion but always lead to the stable reduced order models for stable systems. Bosley and Lees [5] and others have proposed a method of reduction based on the fitting of the time moments of the system and its reduced model but these methods have a serious disadvantage that the reduced order model may be unstable even though the original high order system is stable. Some combined methods are also given for example Shamash [6], Chen, Chang and Han [7] and Wan [8] in which the denominator of the reduced order model is derived by some stability criterion method while the numerator of the reduced model is obtained by some other methods. [9].

In this paper, a new model order reduction method is proposed and its helps in finding the current controller gain value. Simulation results were shows the validity of the proposed method. The proposed model order reduction method is a mixed method, where the numerator polynomial of reduced order model is obtained by using the stability equation method and numerator polynomial is obtained by the method proposed in the paper [10].

## II. DC MOTOR DRIVE

The control schematic of a two-quadrant converter-controlled separately-excited DC motor drive is shown in figure 1. The motor drive shown is a speed controller system. The thyristor bridge converter gets its ac supply through a three phase transformer and fast acting ac contactors. The dc output is fed to the armature of the dc motor. The field is separately excited, and the filed supply can be kept constant or regulated, depending on the need for the field weakening mode of operation. The DC motor has a tachogenerator whose output is utilized for closing the speed loop. The motor is driving a load considered to be frictional for this treatment. The output of the tachogenerator is filtered to remove the ripples to provide the signal, $\omega_{mr}$. The speed command $\omega_r^*$ is compared to the speed signal to produce a speed error signal. This signal is processed through a proportional-plus-integral (PI) controller to determine the torque command. The torque command is limited, to keep within the safe current limits and the current command is obtained by proper scaling. The armature current command $i_a^*$ is compared to the actual armature command $i_a$ to have a zero current error. The PI controller produces the equivalent control signal $V_c$ when an error signal is occurred. The control signal accordingly modifies the triggering angle α to be sent to the converter for implementation.

The current control loop of DC motor drive is shown in the figure 2. The DC machine contains an inner loop due to the induced emf. It is not physically seen; it is magnetically coupled. The inner current loop will cross this back-emf loop, creating a complexity in the development of the model. The inner current loop assures a fast current response and also limits the current to a safer level. The inner current loop makes the converter a linear current amplifier. The outer speed loop ensures that the actual speed is always equal to the commanded speed and that any transient is overcome with in the shortest feasible time without exceeding the motor and converter capability.

The operation of the closed loop speed controlled drive is explained from one or two particular instances of speed command. A speed from zero to rated value is recommended, and the motor is assumed to be at standstill. This will generate a large speed error and a torque command and in turn an armature current command. The armature current error will generate the triggering angle to supply a a preset maximum DC voltage across the motor terminals. The inner current loop will maintains the level permitted by its commanded value, producing a corresponding torque. As motor starts running, the torque and current are maintained at their maximum level, thus accelerating the motor rapidly.

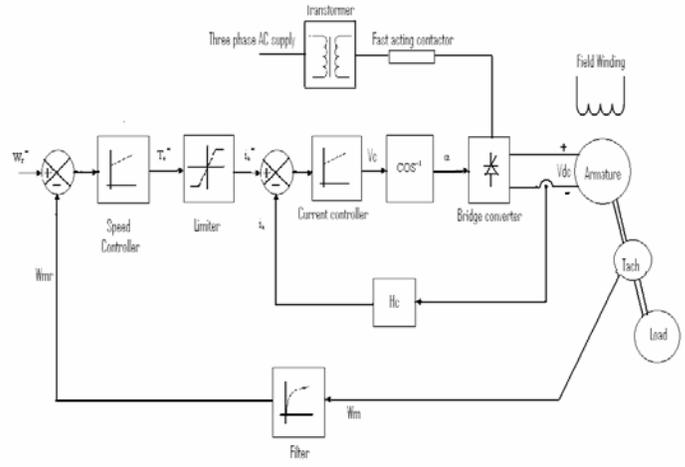

Fig.1. Speed-controlled two-quadrant DC motor drive

When the motor attains the commanded value, the torque command will settle down to a value equal to the sum of the load torque and other losses to keep the motor in steady state. The DC machine contains an inner loop due to the induced emf. It is not physically seen; it is magnetically coupled. The inner current loop cross this back-emf loop, creating a complexity in the development of the model and is shown in the fig.2.

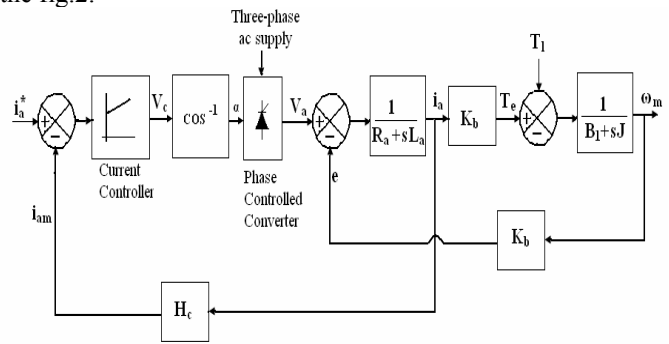

Fig.2. DC motor and current-control loop

The design of the gain constants of the current controllers is of paramount importance in meeting the dynamic specifications of the motor drive.

## III. DESIGN OF CONTROLLERS

The overall closed-loop system of DC motor drive is shown in fig.3. The design of control loops starts from the innermost (fastest) loop to the outer (slowest) loop. The reason to proceed from the inner to the outer loop in the design process is that the gain and time constants of only one controller at a time are solved, instead of solving for the gain and time constants of all controllers simultaneously.

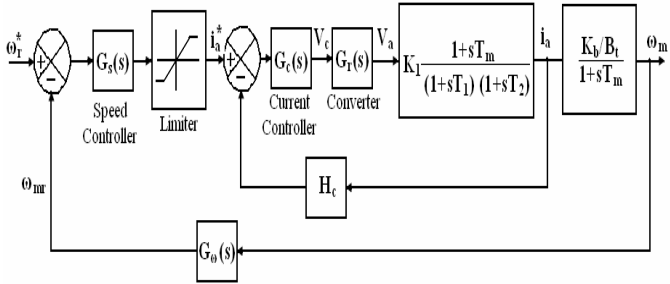

Fig.3. Block diagram of the DC motor drive

In some applications such as traction application, the motor drive need not be speed controlled but may be torque controlled. In that case, the current loop is essential and exists regardless of whether the speed control loop is going to be closed. The performance of the outer loop is dependent on the inner loop; therefore, the tuning of the inner loop has to precede the design and tuning of the outer loop.

## IV. CURRENT CONTROLLER

The current control loop is shown in figure 4.

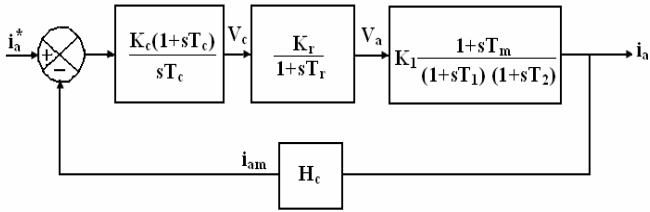

Fig.4. Current-control loop

The loop gain transfer function is,

$$GH_i(s) = \{\frac{K_1 K_C K_r H_C}{T_C}\} \cdot \frac{(1+sT_c)(1+sT_m)}{s(1+sT_1)(1+sT_2)(1+sT_r)} \quad (1)$$

Where,
  $T_1, T_2$ = Electrical time constant of the motor,
  $B_t$ = Total friction co-efficient in N-M/ (rad/sec)
  $J$ = Moment of inertia, kg-m2
  $R_a$ = DC machine armature resistance
  $L_a$ = DC machine armature inductance
  $K_c$ = Current controller gain
  $T_c$ = Current controller time constant
  $K_r$ = Converter gain
  $T_r$ = Converter delay time
  $\omega_m$ = DC motor speed.

This is a fourth order system, and simplification is necessary to synthesize a controller without resorting to a computer. Noting that $T_m$ is on the order of a second and in the vicinity of the gain crossover frequency, we see that the following approximation is valid.

$$(1+sT_m) \cong sT_m \quad (2)$$

This reduces the loop gain function to

$$GH_i(s) = \frac{K(1+sT_c)}{(1+sT_1)(1+sT_2)(1+sT_r)} \quad (3)$$

Where,

$$K = \frac{K_1 K_C K_r H_C T_m}{T_C} \quad (4)$$

The time constants in the denominator are seen to have the relationship

$$T_r < T_2 < T_1 \quad (5)$$

The equation (3) can be reduced to second order, to facilitate a simple controller synthesis, by judiciously selecting

$$T_c = T_2 \quad (6)$$

Then the loop function is

$$GH_i(s) \cong \frac{K}{(1+sT_1)(1+sT_r)} \quad (7)$$

The conventional design methods uses the approximations used in the equations from (2)-(7). We can't assure that the approximations were made are good in the design aspect. So the proposed Model order reduction technique is applied to get the equivalent reduced order model of order two. The validity of the proposed method is verified through the time and frequency domain analysis.

The characteristic equation or denominator of the transfer function between the armature current and its command is

$$(1+sT_1)(1+sT_r)+K=0 \quad (8)$$

This equation is expressed in standard form as

$$T_1 T_r \left( s^2 + s\left(\frac{T_1+T_r}{T_1 T_r}\right) + \frac{K+1}{T_1 T_r}\right) = 0 \quad (9)$$

The general second order characteristic equation is given by,

$$s^2 + 2\xi\omega_n s + \omega_n^2 = 0 \quad (10)$$

On comparing the equations (9) and (10) with the damping ratio $\xi$ as 0.707 for satisfactory operation of system, we get the value for the term K. From that value, the current controller gain $K_c$ can be easily estimated.

## V. MODEL ORDER REDUCTION METHOD

Let the nth order system is given by the transfer function

$$G(s) = \frac{\sum_{i=0}^{n-1} d_i s^i}{\sum_{j=0}^{n} e_j s^j} = \frac{N(s)}{D(s)} = \frac{a_0 + a_1 s + a_2 s^2 + \ldots + a_m s^m}{b_0 + b_1 s + b_2 s^2 + \ldots + b_{n-1} s^{n-1} + b_n s^n} \quad (11)$$

and the corresponding $r^{th}$ (r<n) order reduced order is of the form

$$G_r(s) = \frac{\overline{N(s)}}{\overline{D(s)}} = \frac{d_0 + d_1 s + d_2 s^2 + \ldots + d_{r-1} s^{r-1}}{e_0 + e_1 s + e_2 s^2 + \ldots + e_{r-1} s^{r-1} + e_r s^r} \quad (12)$$

The proposed model order reduction method consists of three steps

Step 1: *The denominator polynomial of reduced order model is obtained by using the stability equation method [11].*

For stable G(s) the eve and odd parts of D(s) may be factored as the following stability equations.

$$D_e(s) = e_0 \prod_{i=1}^{n/2} \left(1 + \frac{s^2}{z_i^2}\right) \quad (13)$$

$$D_o(s) = e_1 \prod_{i=1}^{(n-1)/2} \left(1 + \frac{s^2}{p_i^2}\right) \quad (14)$$

Where,

$$z_1^2 < p_1^2 < z_2^2 < p_2^2 < \ldots\ldots\ldots\ldots$$

After discarding the factors with larger magnitude of zi and pi, the stability equations are reduced, which are combined to give a reduced polynomial of order r, as:

$$D_r(s) = b_0 \prod_{i=1}^{r/2}\left(1+\frac{s^2}{z_i^2}\right) + b_1 \prod_{i=1}^{(r-1)/2}\left(1+\frac{s^2}{p_i^2}\right) = \sum_{j=0}^{r} b_j s^j \quad (15)$$

Step 2: *The numerator polynomial of reduce order system is obtained by using the method proposed in* [10].

Consider the given system transfer function given in (11)

$$G(s) = \frac{a_0 + a_1 s + a_2 s^2 + \ldots + a_m s^m}{b_0 + b_1 s + b_2 s^2 + \ldots + b_{n-1} s^{n-1} + b_n s^n}$$

$$= K \frac{(1 + A_1 s + A_2 s^2 + \ldots + A_m s^m)}{(1 + B_1 s + B_2 s^2 + \ldots + B_{n-1} s^{n-1} + B_n s^n)} \quad ; n \geq m \quad (16)$$

Where, $K = \dfrac{a_0}{b_0}$,

$A_i = \dfrac{a_i}{a_0}$ ; i = 0,1,2,3…m and

$B_j = \dfrac{b_j}{b_0}$ ; j = 0,1,2,3…n

Let the transfer function of the approximating low-order system be,

$$G_r(s) = K \frac{(1 + C_1 s + C_2 s^2 + \ldots + C_q s^q)}{(1 + D_1 s + D_2 s^2 + \ldots + D_{p-1} s^{p-1} + D_p s^p)} \quad (17)$$

where n≥p≥q

The coefficients of $D_1$, $D_2$, .. were obtained from the equation (15).

The following relation should be satisfied as closely as possible

$$\frac{|G(j\omega)|^2}{|G_r(j\omega)|^2} = 1 \text{ for } 0 \leq \omega \leq \infty \quad (18)$$

$$\frac{G(s)}{G_r(s)} = \frac{(1+A_1s+A_2s^2+\ldots+A_ms^m)(1+D_1s+D_2s^2+\ldots+D_{p-1}s^{p-1}+D_ps^p)}{(1+B_1s+B_2s^2+\ldots+B_{n-1}s^{n-1}+B_ns^n)(1+C_1s+C_2s^2+\ldots+C_qs^q)}$$

$$= \frac{(1 + m_1 s + m_2 s^2 + \ldots + m_u s^u)}{(1 + l_1 s + l_2 s^2 + \ldots + l_{v-1} s^{v-1} + l_v s^v)} \quad (19)$$

Where, u=m+p and v=n+q

$$\frac{|G(j\omega)|^2}{|G_r(j\omega)|^2} = \left.\frac{G(s)G(-s)}{G_r(s)G_r(-s)}\right|_{s=j\omega} \quad (20)$$

The equation produces the even powers of 's' and can be written as,

$$\frac{|G(j\omega)|^2}{|G_r(j\omega)|^2} = \left.\frac{1 + L_2 s^2 + L_4 s^4 + \ldots + L_{2u} s^{2u}}{1 + M_2 s^2 + M_4 s^4 + \ldots + M_{2v} s^{2v}}\right|_{s=j\omega} \quad (21)$$

$$\frac{|G(j\omega)|^2}{|G_r(j\omega)|^2} = 1 + \frac{(L_2-M_2)s^2 + (L_4-M_4)s^4 + \ldots + (L_{2u}-M_{2v})s^{2u}}{1+M_2s^2+M_4s^4+\ldots+M_{2v}s^{2v}}\bigg|_{s=j\omega} \quad (22)$$

To satisfy the above equation,
$L_2 = M_2$
$L_4 = M_4$
.
.
.
$L_{2u} = M_{2v}$     ; if u=v     (23)

If u<v, then the error generated by the lower order model is,

$$|\varepsilon| = \frac{|G(j\omega)|^2}{|G_r(j\omega)|^2} - 1 \quad (24)$$

From the above equations, we can obtain the conditions as

$$L_{2x} = \sum_{i=0}^{x-1}(-1)^i 2m_i m_{2x-i} + (-1)^x m_x^2 \quad (25)$$

for x= 1,2, 3……..u and $m_0$=1

and

$$M_{2y} = \sum_{i=0}^{y-1}(-1)^i 2l_i l_{2y-i} + (-1)^x l_y^2 \quad (26)$$

for y= 1,2, 3……..v and $l_0$=1

From the equation (11), the reduced order model numerator coefficients can be obtained. Finally the reduced order model is in the form of

$$Gr(s) = K \frac{(1+C_1s+C_2s^2+\ldots+C_qs^q)}{(1+D_1s+D_2s^2+\ldots+D_{p-1}s^{p-1}+D_ps^p)} \quad (27)$$

Step 3: *The coefficients of the reduced order denominator polynomial is adjusted further to get the better approximation with original system. The coefficient of 's' term in the denominator is increased by n% and the same was reduced from the coefficient of the term 's²'. The value of n is choosen by trial and error method. Normally the value of n is ranging between 1 to15.*

VI. DESIGN EXAMPLE

The motor parameters and ratings of a speed controlled DC motor drive maintaining the field flux constant are as follows 220V, 8.3A, 1470 rpm, $R_a$=4Ω, J=0.0607kg-m², $L_a$=0.072H, $B_t$=0.0869N-m/rad/sec, $K_b$=1.26V/rad/sec. The converter is supplied from 230V, 3-PhaseAC at 50Hz. The converter is linear, and its maximum control input voltage is ±10V. The tachogenerator has the transfer function $G_\omega(s) = \frac{0.065}{1+0.002s}$. The speed reference voltage has a maximum of 10V. The maximum current permitted in the motor is 20A.

*Converter transfer function:*

$$K_c = \frac{1.35V}{V_{cm}} = \frac{1.35 \times 230}{10} = 31.05$$

$V_{dc}$(max) = 310.5 V

The rated DC voltage required is 220V, which corresponds to a control voltage of 7.09V. The transfer function of the converter is,

$$G_r(s) = \frac{31.05}{(1+0.00138s)} \quad (28)$$

*Current transducer gain:* The maximum safe control voltage is 7.09V, and this has to correspond to the maximum current error:

$i_{max}$ = 20A

$$H_c = \frac{7.09}{I_{max}} = 0.355 V/A$$

*Motor transfer function:*

$$K_1 = \frac{B_t}{K_b^2 + R_a B_t} = 0.0449$$

T1=0.1077 sec and T2=0.0208 sec

$$T_m = \frac{J}{B_t} = 0.7 \sec$$

The subsystem transfer functions are

$$\frac{I_a(s)}{V_a(s)} = K_1 \cdot \frac{(1+sT_m)}{(1+sT_1)(1+sT_2)} = \frac{0.0449(1+0.7s)}{(1+0.0208s)(1+0.1077s)} \quad (29)$$

$$\frac{\omega_m(s)}{I_a(s)} = \frac{K_b/B_t}{(1+sT_m)} = \frac{14.5}{(1+0.7s)} \quad (30)$$

*Design of current controller:*

$T_2$ = 0.0208 sec; $T_c$=0.03 sec

By applying the proposed model order reduction technique, with the value of ζ=0.707, the value of K is obtained as,

K= 357.192

$$K_C = \frac{KT_C}{K_1 H_C K_r T_m} = 35.719$$

By using trial and error method the value of Kc can be adjusted in to the value of 3.1 to obtain the better response for the current controller. The validity of this proposed method is evaluated by plotting the frequency response of the closed loop current to its command. This is shown in fig.5.

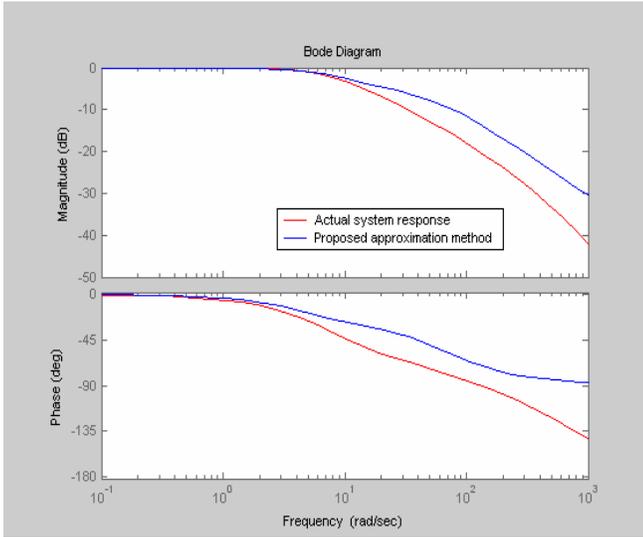

Fig.5 Frequency responses of actual and reduced order systems

From this figure, it is evident that the proposed method is quite valid in the frequency range of interest. The same in time domain analysis is shown in fig.6.

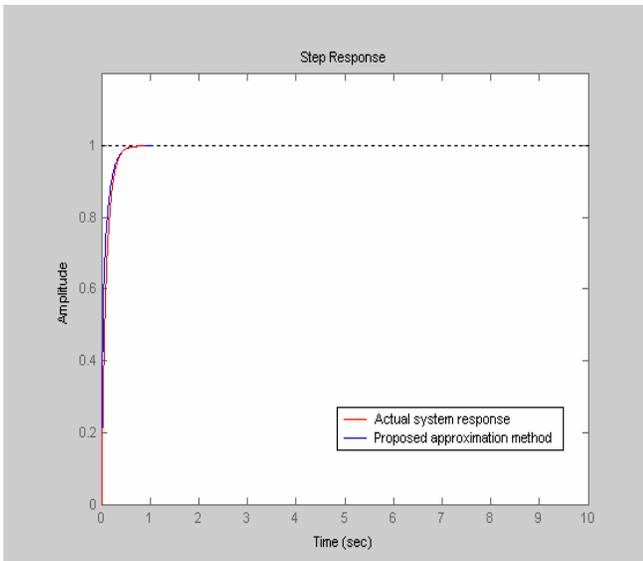

Fig.6 Time responses of actual and reduced order systems

The time responses are important to verify the design of the controller and it is shown in fig. 7 for different values of current controller gain, $K_c$.

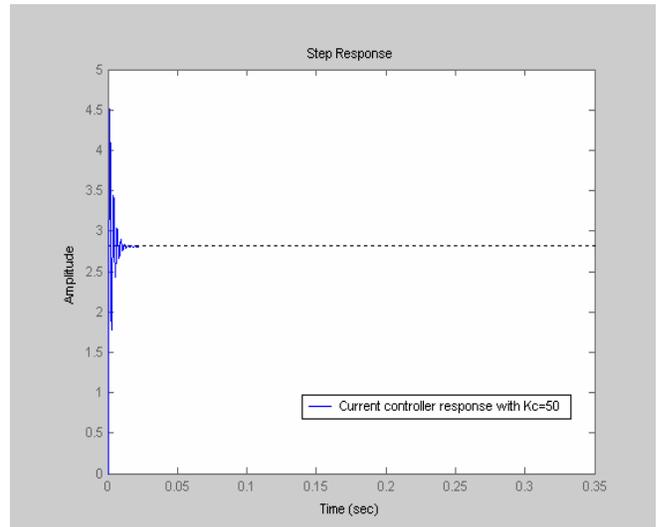

(a)

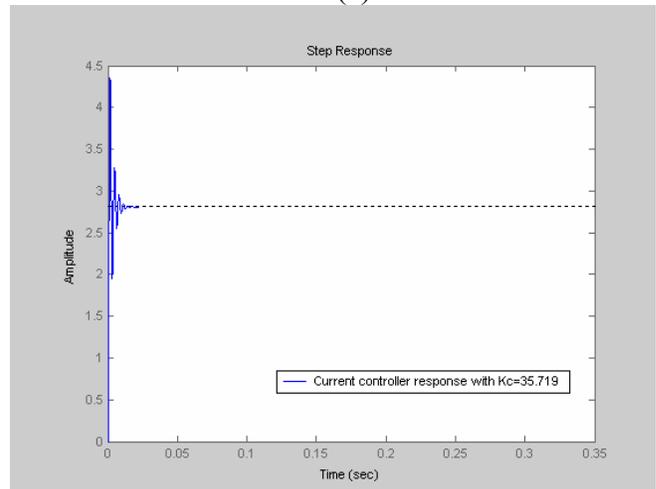

(b)

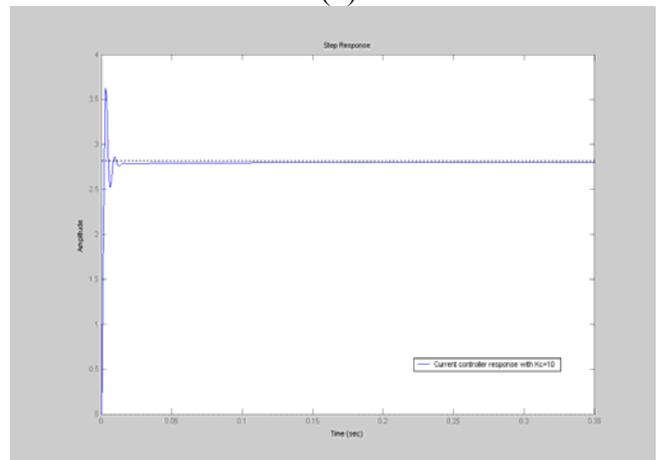

(c)

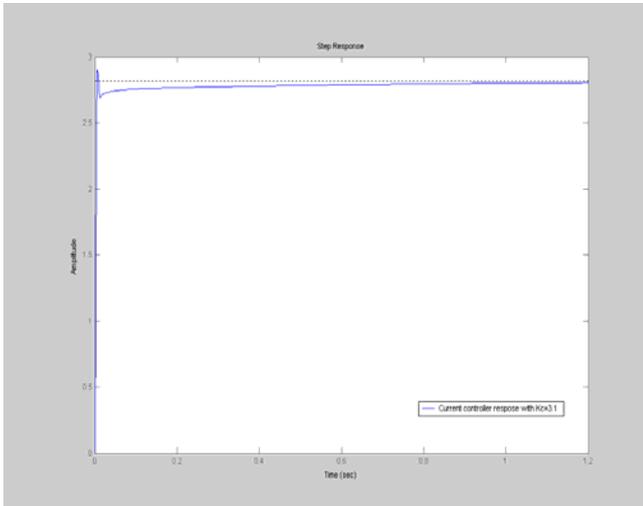

(d)

Fig.7 (a) Current controller response with $K_c=50$ (b) Current controller response with $K_c=35.719$ (c) Current controller response with $K_c=10$ (d) Current controller response with $K_c=3.1$

The value of $K_c$ can be optimized by trail and error method. The integral square error (ISE) was calculated in between the actual and approximated system as 0.0204 and this shows the validity of the proposed method.

## VII. CONCLUSION

The design of the gain constant of current controller is of paramount importance in meeting the dynamic specifications of the motor drive. This paper helps to obtain the same with the help of model order reduction technique. The validity of the proposed method was illustrated through the design example. The proposed model order reduction method is mathematically simple and produces the stable reduced order system if the given system is stable. The optimal value of current controller gain to suppress the oscillations at the output was obtained by trail and error method but it may be selected optimally by applying the neural network concepts or genetic algorithm.

### AUTHORS PROFILE

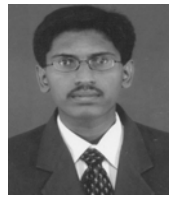

**Ramesh.K** received B.E. degree in Electrical and Electronics Engineering from K.S.R.College of Technology, Tamilnadu, India in 2002 and M.E. in Applied Electronics from Kongu Engg. College, Anna University, Chennai in 2005 and also received the MBA in System from PRIDE, Salem (Tamilnadu), India in 2005. Since then, he is working as a Assistant Professor in Velalar College of Engineering and Technology (Tamilnadu), India. Presently he is a Part time (external) Research Scholar in the Department of Information and Communication at Anna University, Chennai (India). His fields of interests include Model order reduction, Controller design and Optimization Techniques.

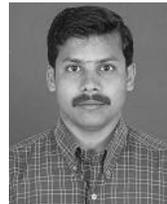

**Ayyar.K** received B.E. Degree in Electrical and Electronics Engineering from Ranipettai Engineering College, Tamilnadu, India in 2003 and M.E. in Power Electronics and Drives from Government College of Engineering, Salem, Anna University, Chennai in 2007. Since then, he is working as a Lecturer in Velalar College of Engineering and Technology (Tamilnadu), India. Presently he is a Part time (Internal) Research Scholar in the Department of Electrical and Electronics Engineering at Anna University, Coimbatore (India). His fields of interests include Power Drives and control, Controller design and System Optimization.

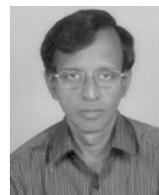

**Dr.Nirmalkumar.A**, received the B.Sc.(Engg.) degree from NSS College of Engineering, Palakkad in 1972, M.Sc.(Engg.) degree from Kerala University in 1975 and completed his Ph.D. degree from PSG Tech in 1992. Currently, he is working as a Professor and Head of the Department of Electrical and Electronics Engineering in Bannari Amman Institute of Technology, Sathyamangalam, Tamilnadu, India.

His fields of Interest are Power quality, Power drives and control and System optimization.

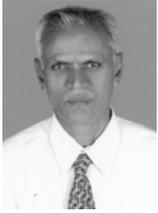

**Dr.G.Gurusamy** obtained his Pre-University education at St Johns College, Palayamkottai, and Thirunelveli district. He then joined PSG College of Technology, Coimbatore, in the year 1962 to pursue his Engineering course. He was graduated in Electrical Engineering in 1967. He latter obtained M.E (Applied Electronics) in 1972 and Ph.D in Control Systems in 1983. He is currently working as a Dean of Department of Electrical and Electronics Engineering in Bannari Amman Institute of Technology, Sathyamangalam. His fields of Interest are Advanced control, Digital control, Optimization and Biomedical electronics.